\documentclass{ws-ijmpd}

\def\rfr#1{Eq.~(\ref{#1})}

\def\bm#1{{\mbox{\boldmath$#1$\unboldmath}}}

\def\bar{\begin{eqnarray}}
\def\ear{\end{eqnarray}}

\def\eqi{\begin{equation}}
\def\eqf{\end{equation}}
\def\eqia{\begin{eqnarray}}
\def\eqfa{\end{eqnarray}}
\def\rp#1#2{{#1\over#2}}

\def\lb#1{\label{#1}}

\def\oc2{$\mathcal{O}(c^{-2})$}

\begin{document}

\markboth{L. Iorio, V. Lainey}
{The Lense-Thirring effect in the Jovian system}

%
\def\draftnote{}
%

\title{THE LENSE-THIRRING EFFECT IN THE JOVIAN
SYSTEM OF THE GALILEAN SATELLITES AND ITS MEASURABILITY}

\author{LORENZO IORIO}

\address{Viale Unit$\grave{a}$ di Italia 68,\\
Bari, 70125,
Italy\\
lorenzo.iorio@libero.it}

\author{VAL$\acute{\rm E}$RY LAINEY}

\address{Observatoire Royal de Belgique, Avenue Circulaire 3, Bruxelles 1180,
Belgium and IMCCE/Observatoire de Paris, UMR 8028 du CNRS,
77 Avenue Denfert-Rochereau, F-75014, France\\
Lainey@oma.be}

\maketitle

\begin{history}
\received{Day Month Year}
\revised{Day Month Year}
\comby{Managing Editor}
\end{history}

\begin{abstract}
In this paper we investigate the possibility of measuring the
post-Newtonian general relativistic gravitomagnetic Lense-Thirring
effect in the Jovian system of its Galilean satellites Io, Europa,
Ganymede and Callisto in view of recent developments in processing and modelling their optical observations spanning a large time interval (125 years). The present day best
observations have an accuracy between several kilometers to few tens
of kilometers, which is just the order of magnitude of the Lense-Thirring shifts of the orbits
of the Galilean satellites over almost a century. From a comparison between analytical development
and numerical integration it turns out that, unfortunately, most of the secular component of the gravitomagnetic signature
is removed in the process of fitting the initial conditions. Indeed, an estimation of the
magnitude of the Lense-Thirring effect in the ephemerides residuals is
given; the resulting residuals have a maximum
magnitude of 20 meters only (over 125 years). 
\end{abstract}


\section{Introduction}

\subsection{The post-Newtonian gravitomagnetic Lense-Thirring effect}
One of the few predictions of the Einstein's General Theory of
Relativity (GTR) which is still awaiting for a direct and reliable
observational check is the so called Lense-Thirring (LT) effect\cite{lenti}. 
It is a consequence of the off-diagonal components
$g_{0i},i=1,2,3$ of the space-time metric generated by a weakly
gravitating and slowly rotating massive body of mass $M$ and
proper angular momentum $\bm{S}$ in the linearized weak-field and
slow-motion approximation of GTR\cite{mash,tart}. The mass-energy currents induce,
among other things, a Lorentz-like, velocity-dependent force which
acts on the orbital motion of a test particle freely moving in the
gravitational field of the central body. Such force is
\eqi\bm{F}=-2m\left(\rp{\bm{v}}{c}\right)\times\bm{B}_g,\lb{forza}\eqf
where $\bm v$ is the particle's velocity and \eqi
\bm{B}_g=\rp{G[3\bm r(\bm r\cdot \bm S )-r^2 \bm
S]}{cr^5}.\lb{gmfield}\eqf $G$ is the Newtonian gravitational
constant and $c$ is the speed of light in vacuum. Due to the
formal resemblance of $\bm B_g$ with the dipolar magnetic field of
the Maxwellian electromagnetism, the ensemble of gravitational
phenomena induced by the mass currents is named gravitomagnetism\cite{mash}. 
It turns out that the non-central gravitomagnetic
force of \rfr{forza} makes the longitude of the ascending node
$\Omega$ and the argument of pericentre $\omega$ to undergo tiny
secular precessions\cite{lenti,Soffel 1989}\cdash\cite{Iorio 2001}
\eqi\rp{d\Omega_{\rm LT}}{dt}=\rp{2GS}{c^2 a^3(1-e^2)^{3/2}},\
\rp{d\omega_{\rm LT}}{dt}=-\rp{6GS\cos i}{c^2
a^3(1-e^2)^{3/2}},\lb{leti}\eqf where $a,e,i$ are the semimajor
axis, the eccentricity and the inclination yto the equator of the orbit of the
moving particle, respectively.

The gravitomagnetic phenomena may have strong consequences in many
astrophysical and astronomical scenarios involving, e.g.,
accreting disks around black holes\cite{Stella et al. 2003},
gravitational lensing and time delay\cite{Sereno 2003,Sereno 2005}. 
Unfortunately, in these contexts the knowledge of the
various competing effects is rather poor and makes very difficult
to reliably extract the genuine gravitomagnetic signal from the
noisy background. E.g., attempts to measure the LT effect around
black holes are often confounded by the complexities of the
dynamics of the hot gas in their accretion disks. Conversely, if
one believes in GTR, the LT effect could be used, in principle,
for measuring the proper angular momentum of the central mass
which acts as sources of the gravitomagnetic field.

It must be noted that, according to K. Nordtvedt, gravitomagnetism
would have already been indirectly tested  from the radial motion
of the Earth's geodetic satellite LAGEOS\cite{nor88} and from the
high-precision reconstruction of the lunar orbit with the Lunar
Laser Ranging (LLR) technique\cite{nor03}.

An attempt to directly measure the LT effect in the gravitational
field of the Earth has been performed by Ciufolini and Pavlis with the LAGEOS and LAGEOS
II satellites\cite{ciu04} by using an observable proposed by one of us in Ref.~\refcite{Iorio and Morea 2004}. 
The claimed total accuracy is
5-10$\%$ at 1-3 sigma, but such estimates have been criticized by
one of us in Ref.~\refcite{Iorio 2005a} who proposes a 19-24$\%$ total error at
1-sigma. The main limiting factors are the systematic errors of
gravitational origin due to the static and secularly varying parts
of the even zonal coefficients $J_{\ell}$ of the multipolar
expansion of the terrestrial gravitational potential.

The first, very preliminary, evidence of the gravitomagnetic field of
the Sun through the LT precessions of the longitudes of the
perihelia $\varpi$ of the inner planets of the Solar System has
recently been reported by one if us in Ref.~\refcite{Iorio 2005b}.

In April 2004 the GP-B spacecraft\cite{Everitt et al 2001} has been
launched. Its aim is the measurement of another gravitomagnetic
effect, i.e. the precession of the spins\cite{Schiff 1960} of
four superconducting gyroscopes carried onboard with an expected
accuracy of $1\%$ or better.

\subsection{Aim of the paper}
In this paper we want to investigate the possibility of measuring
the LT effect on the orbits of the Galilean satellites Io (1),
Europa (2), Ganymede (3) and Callisto (4) evolving in the gravitational
field of Jupiter in view of recent improvements in their
ephemerides\cite{Lainey et al 2004a,2004b}. The orbital
parameters of the Galilean satellites of Jupiter are presented in Table
\ref{galpar}.
\begin{table}[ph]
  \tbl{Averaged Keplerian orbital elements of the Galilean satellites Io, Europa, Ganymede and Callisto deduced from L1 ephemerides.}
 {\begin{tabular}{@{}cccc@{}} \toprule
Satellite & $a$ (km)& $e$ & $i$ (deg)\\ \colrule
Io & \hphantom{0}422,030 &  0.0042 &  0.036\\
Europa & \hphantom{0}671,261 & 0.0094 & 0.469\\
Ganymede & 1,070,621 & 0.0015 & 0.175\\
Callisto & 1,883,134 &  0.0075 & 0.187\hphantom{0}\\ \botrule
\end{tabular} \label{galpar}}
\end{table}
It is noteworthy that the idea of using the Jovian system was put
forth for the first time by Lense and Thirring themselves in their
original paper\cite{lenti}.

\section{The R-T-N scheme for the Lense-Thirring effect}
Since the Galilean satellites of Jupiter move along nearly
circular and equatorial orbits and to make easier the comparison
with the current ephemerides we will consider the radial,
transverse and normal components of their orbits. Their
perturbations can be expressed in terms of the integrated shifts
of the Keplerian orbital elements $\Delta a,\Delta e,\Delta
i,\Delta\Omega,\Delta\omega,\Delta\mathcal{M}$, where
$\mathcal{M}$ is the mean anomaly, as\cite{chri88}
\eqi\Delta R=\sqrt{(\Delta a)^2+\rp{\left[(e\Delta a +a\Delta e
)^2+(ae\Delta{\mathcal{M}})^2\right]}{2}},\lb{radi}\eqf
\eqi\Delta
T=a\sqrt{1+\rp{e^2}{2}}\left[\Delta{\mathcal{M}}+\Delta\omega+\cos
i\Delta\Omega +\sqrt{(\Delta e)^2+(e\Delta
{\mathcal{M}})^2}\right],\lb{tranv}\eqf
\eqi\Delta N=a\sqrt{\left(1+\rp{e^2}{2}\right)\left[\rp{(\Delta i
)^2}{2}+(\sin i\Delta\Omega)^2\right]}.\lb{norm}\eqf

>From \rfr{leti} it turns out that the gravitomagnetic force only
affects, in general, both the transverse and the normal
components: $\Delta R_{\rm LT}=0$. For polar orbital
configurations, i.e. $i=90$ deg, $\Delta T_{\rm LT}=0$, while for
equatorial orbits, i.e. $i=0$ deg, $\Delta N_{\rm LT}=0$.
For nearly circular and equatorial orbits it turns out that
$\Delta T\sim a\Delta\lambda$, where
$\lambda=\mathcal{M}+\Omega+\omega$ is the mean longitude.

\section{The Lense-Thirring effect in the Jovian system}
The four Galilean satellites\footnote{For the physical parameters
of Jupiter and its moons see
http://ssd.jpl.nasa.gov/sat$\_$gravity.html} of Jupiter are
currently the best candidates (among satellites evolving around
giant planets) in regard to their orbital motion's accuracy.
Indeed, ephemerides of the Galilean satellites reach an accuracy
of only few tens of kilometers for the modern period\cite{2004b},
and benefit of observations dispatched over more than a century
(from 1891 to 2003 in the case of L1 ephemerides\footnote{These
ephemerides are available at {http://www.imcce.fr}.}).
These observations have various origin. While most of them are
photographic ones, the observation of mutual events
(the eclipse or occultation of one satellite by another) delivers
the most accurate Earth-based observations of the Galilean system.
Indeed, such observations are photometric instead of astrometric,
and so are less sensitive to atmosphere turbulence.
Hence, the accuracy of these latter can reach few tens of kilometers.

More, JPL ephemerides accuracy amount to 5 km for the present period
by the use of unpublished spacecraft observations (see Ref.~\refcite{JUP230}
and http://ssd.jpl.nasa.gov/sat$\_$eph.html).


\subsection{Theoretical predictions: analytical and numerical estimates}\label{sec:tpred}
In regard to the LT effect, for the Jupiter's angular momentum we
assume $S=4.33\times 10^{38}$ kg m$^2$ s$^{-1}$ since the
ratio\footnote{It amounts to $2/3\sim 0.6$ for a hollow spherical
shell and to $2/5=0.4$ for a homogenous sphere.} $\alpha$ of the
moment of inertia $I$ to $MR^2$, where $R$ is the mean equatorial
radius, amounts to\cite{Irwin 2003} $0.264$
(see also http://nssdc.gsfc.nasa.gov/planetary/factsheet/jupiterfact.html).
Such estimates are based on theoretical models: at present, there
are no direct, independent measurements of $S$ for Jupiter. It
should also be noted that Lense and Thirring in Ref.~\refcite{lenti} and
Soffel in Ref.~\refcite{Soffel 1989} overestimated the gravitomagnetic
precessions of the Galilean satellites because they modeled
Jupiter as a uniform sphere by assuming $\alpha=0.4$.

In order to compute the secular LT shifts $\Delta T_{\rm LT}$ on
the Jovian Galilean moons it is important to consider that such a
system is very complex because of the strong mutual perturbations
of one satellite on another. In particular, the Laplacian
resonance\cite{Duriez 1982,Peale and Lee 2002} \eqi N_1-3N_2+2N_3
=0\lb{reson}\eqf establishes a relation among the Keplerian
mean mean motions\footnote{By defintion, the mean mean motion is the secular component in the mean
longitude expression. Hence, it already contains the node and pericenter
secular variations. So mean mean motion is different from the averaged mean
motion (that contains only the mean anomaly). See Ref.~\refcite{Laskar and Jacobson 1987}.} $N$ of Io, Europa and Ganymede. This has
consequences also on the LT effect. Indeed, the gravitomagnetic
force of \rfr{forza} is a small correction with respect to the
Newtonian monopole and, as usual in perturbation theory, it must
be evaluated onto the unperturbed Keplerian path for which the
velocity entering \rfr{forza} is approximately $v\sim aN$. Thus,
the gravitomagnetic components of the equations of motion for Io,
Europa and Ganymede are not independent but they are coupled due
to \rfr{reson}. More, 
secular variations of nodes and pericentres will  
affect the mean mean motions. To preserve the Laplacian relation, 
variations on mean anomalies are also expected.
Hence, longitude evolutions of Io, Europa and Ganymede have to be studied. 
This means that differences with respect to the
case of independent motions, illustrated in Table \ref{nores}, are
expected. This feature has been investigated numerically in the
following way.
\begin{table}
  \tbl{Secular LT shifts, in km, over 125 years calculated with \rfr{leti} and \rfr{tranv}
  by neglecting the mutual perturbations of one satellite on another. For the Jupiter's angular momentum
  the value $S=4.33\times 10^{38}$ kg m$^2$ s$^{-1}$ has been adopted.}
  {\begin{tabular}{@{}ccccc@{}} \toprule
& Io & Europa & Ganymede & Callisto\\
  \colrule
$\Delta T_{\rm LT}$ (km) & -28 & -11 & -4 & -1\\
\botrule
\end{tabular}\label{nores}}
\end{table}

\subsection{Numerical simulations}
In order to numerically investigate the effect of the Jovian
gravitomagnetic field on the orbital motion of the four Galilean
satellites, we have performed several numerical integrations.
\newline
The adopted software is called NOE (Numerical Orbital Elaboration)
and is inspired from a former work presented in Ref.~\refcite{Lainey et al 2004a}. 
It was developed at the Royal Observatory of Belgium
mainly for natural satellites ephemerides purpose. It is a N-body
code which incorporates highly sensitive modeling and can generate
partial derivatives. The latter ones are needed to fit the initial
positions, velocities and other parameters to the observation
data. We used the exact modeling, initial conditions and parameter values that were 
used during L1 ephemerides elaboration. Let us recall that this model introduces the Jovian gravity 
field by mean of $J_2, J_4, J_6$ coefficients, the satellites mutual perturbations (including their 
respective oblateness coefficients $J_2$ and $c_{22}$) and the Solar perturbation using
DE406 ephemerides\footnote{In particular, the indirect planetary perturbations are implicitely introduced.}.
The selection ot these perturbations was done after a careful study of the magnitude expected from a large set of usually 
neglected perturbations (planetary and satellites precessions, Jovian $J_3$ coefficient ...), but did not considered 
LT effect at that time. A perturbation was added in the model only when found significant from the observations accuracy.
In particular, secular variations inducing less than hundreds of kilometers after one century on the satellite longitudes have 
generally been considered negligible. Indeed, these latter can be easilly absorbed by tiny changes on initial satellite positions 
and velocities. On the other hand, small perturbations may be retained if damping enough known frequencies in the system or 
adding new ones.

LT effect has been introduced by means of \rfr{forza} and
\rfr{gmfield}. For an explicit formulation of all the equations used, 
we refer to Ref.~\refcite{Lainey et al 2004a}. Initial conditions and parameter values 
are available in Ref.~\refcite{2004b}. 
In particular, L1 ephemerides
result from the fit of a high sensitive model to observations
covering a time span from 1891 to 2003. 

The integrator subroutine is the one of Everhart\cite{Evehart 1985}, called RA15. 
It was chosen for its speed and accuracy. During the integration a
constant step size of $\Delta t=0.08$ day was used. To increase
the numerical accuracy during the fit procedure (see subsection
\ref{sec:mes}) we performed integration over $\pm 62.5$ years
instead of $+125$ years. The numerical accuracy of our simulations
is at the level of several meters.

\begin{figure}[pb]
\centerline{\psfig{file=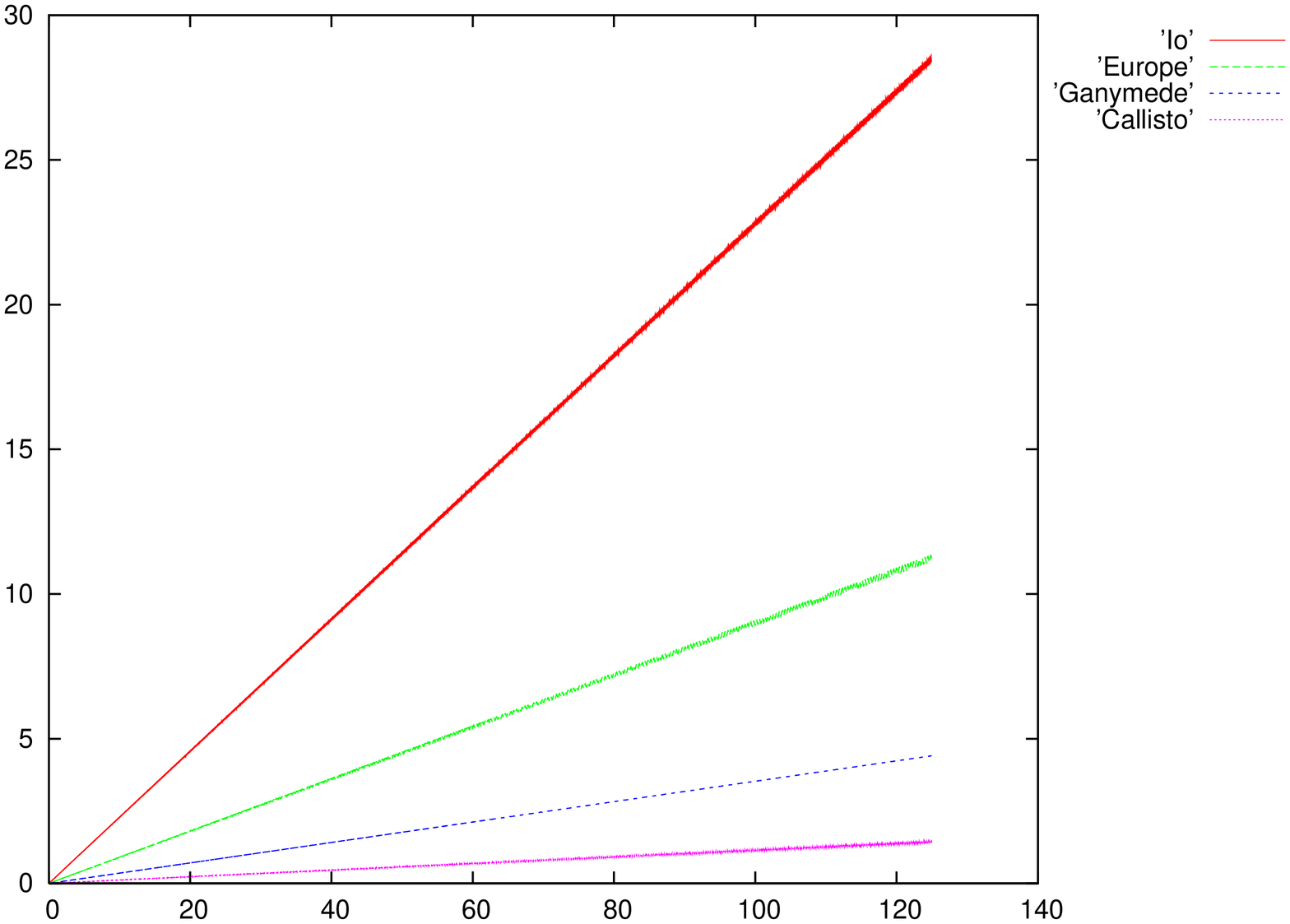,width=4.7cm,angle=-90}\psfig{file=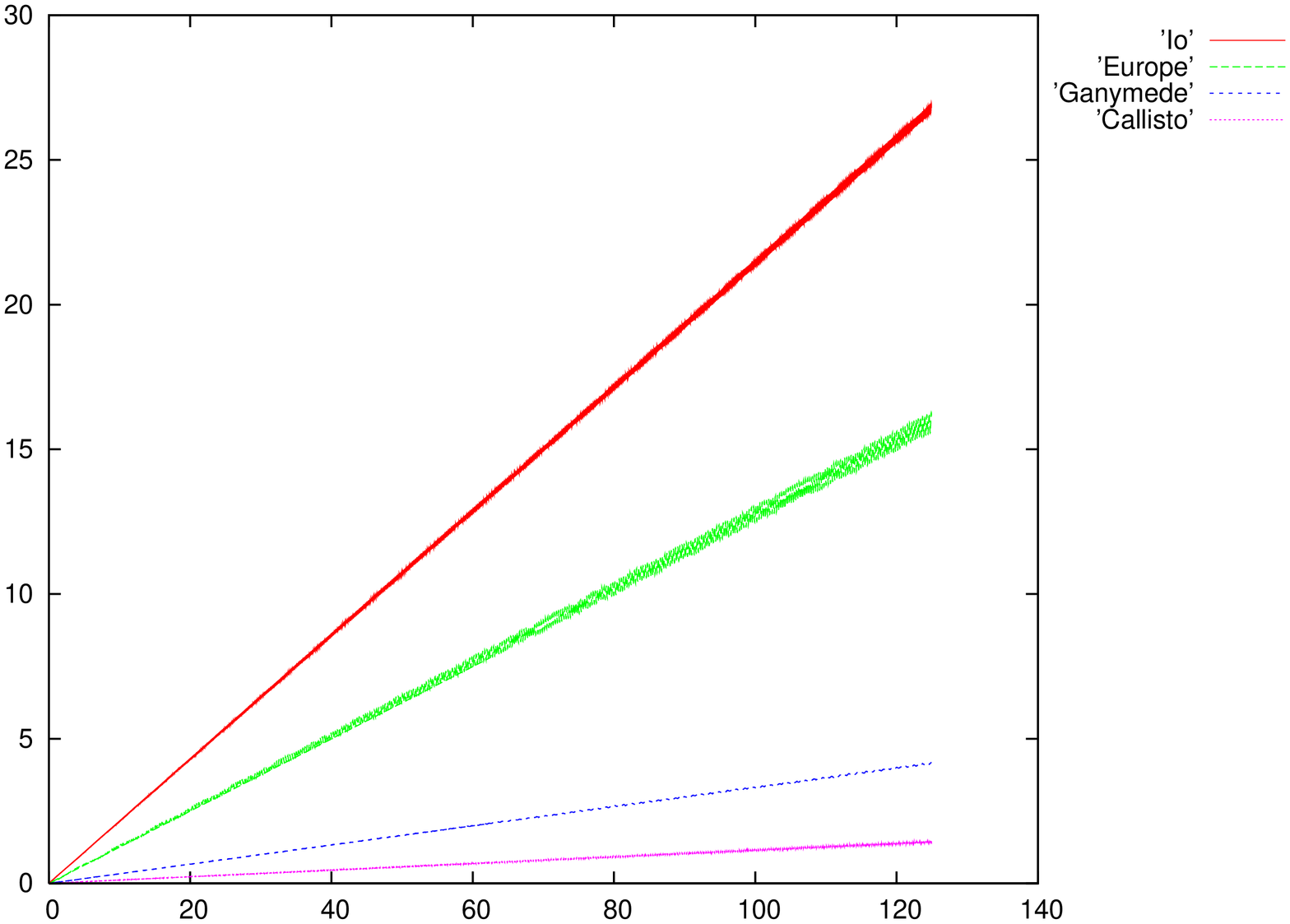,width=4.7cm,angle=-90}}
\vspace*{8pt}
\caption{
Differences in distance for the four
Galilean satellites between a simulation neglecting the LT effect
and a simulation including the LT effect.  The case without the
Laplacian resonance is shown on the left, while it has been included
on the right. The horizontal axes are in years and the vertical
axes are in km. In both cases the initial conditions and parameter values
have been taken from Ref.~19.\label{distnores}
}
\end{figure}

In Fig.~(\ref{distnores}) we plot the differences on satellite distances between a numerical simulation
including LT effect and a second simulation (using the same set of initial conditions)
but neglecting LT effect. The adopted time span is 125 years and covers roughly
the modern observation period. In a first step we followed our former analytical formulation
and so neglected the mutual perturbations (by nullifying the satellite masses). In particular, the
Laplacian resonance was not present in the system at this step. Results are shown in the left panel
of Fig.~(\ref{distnores}).

As expected, the obtained secular shifts agree with the analytical
calculations of Table \ref{nores}.

In a second step, we reintroduced the satellite masses values
(and, thus, the Laplacian resonance). Results are shown in the
right panel of Fig.~(\ref{distnores}). As it can be seen, the
shifts remain unchanged for Callisto at the level of 1.4 km.
Rather small change appear for Ganymede with a slight decrease
from 4.4 km to 4.2 km; Io and Europa experience the most important
changes passing from 28.6 km to 27 km and from 11.3 km to 16.3 km,
respectively. These differences with our numerical estimations are
induced by the Laplacian resonance as explained in Section
\ref{sec:tpred}.

\subsection{Measurability of the Lense-Thirring effects with the current ephemerides}\label{sec:mes}

Galilean satellite observations are rather different from those for the artificial satellites,
in the sense that only the right ascension and declination for each natural satellite are available. 
Neither estimation of the Jupiter-satellite distances nor of satellite velocities are possible. 
Hence, reconstruction of observed Keplerian elements is not possible. Moreover, the observations 
have a quite smaller coverage by satellite revolutions and are not dispatched equally over the years. 
This is why the method used to detect LT effects with LAGEOS satellites cannot be applied here.
However, and to estimate the possible detection of LT effect 
among Galilean system in a realistic way, we decided to apply a fit procedure of our 
numerical simulation neglecting LT effect on the simulation including LT effect (acting like some real observations). 
The induced residuals can then be anticipated as part of the real observed residuals. If these former 
are found signicant (at least one order of magnitude less), then one can deduce that the effect is detectable.

This method was already applied successfully to estimate tidal effects among the Galilean system in Ref.~\cite{LaineyTobie}.


We performed the fit procedure following ephemerides elaboration. As for L1 ephemerides,
we fitted only the initial positions and
velocities of each satellite. A sample of 628 daytimes with a 72.8 days
time step was used. Differences on Cartesian positions for all satellites have been fitted,
with no weights assigned.

\begin{figure}[pb]
\centerline{\psfig{file=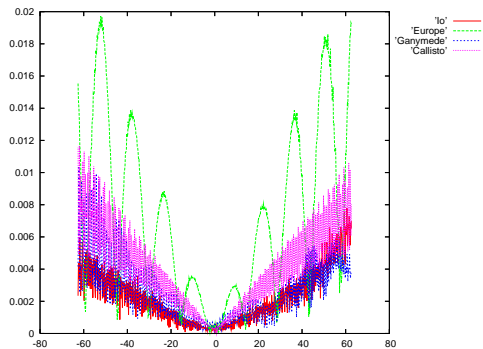,width=4.7cm,angle=-90}}
\vspace*{8pt}
\caption{Differences in distance for the four
Galilean satellites between a simulation neglecting the LT effect
and a simulation including the LT effect. The initial conditions have been fitted. The
horizontal axes are in years and the vertical axes are in km.\label{distres}}
\end{figure}
In Fig.~(\ref{distres}) we show the residuals induced by LT effect
after fit and which account in ephemerides residuals. It can be
noted that a large part of the secular LT signature has been
removed from the differential time series over 125 years to a
level of several meters (the level of accuracy of our
simulations), while an interesting, long period pattern appears
for Europa with an amplitude of 20 meters. Most of this variation
appear on the inclination and the mean longitude (graphs not shown
here).

In Table \ref{ipos} the corrections delivered by our fit procedure
and applied to the initial positions and velocities are given.
Table \ref{ikep} presents the same corrections converted into
Keplerian elements. In the case of Callisto, significant changes
appear only on $\Omega$ and $\omega$ with opposite numerical
values. That is consistent with our analytical approach. For the
three other satellites, significant variations appear both on
$\Omega$ and $\omega$ and $\mathcal{M}$, which is a consequence of
the Laplacian resonance.

\begin{table}
 \tbl{Corrections to the initial values of the satellite
  positions (AU) and velocities (AU/d) in a J2000 Earth mean equatorial frame centered on Jupiter.}
{\begin{tabular}{@{}cccc@{}} \toprule
Satellite & $x\ \ \ \ \ \ \ \ \ \ $ & $y\ \ \ \ \ \ \ \ \ \ $ & $z\ \ \ \ \ \ \ \ \ \ $\\
\colrule
Io's position& -1.890118053390E-012& -4.300639705467E-013& -1.844960816066E-013\\
Io's velocity& -4.719160826005E-012& -3.184161051147E-012& -1.966170498982E-012\\
Europa's position&  4.161540903546E-013&  7.473039114608E-013& -1.666243857299E-012\\
Europa's velocity&  1.710567885254E-012& -1.396029125633E-012&  8.221409664166E-013\\
Ganymede's position &  1.634014971940E-012& -3.859797030603E-012& -8.237275878758E-013\\
Ganymede's velocity&  2.916018121412E-012& -1.678258226833E-013&  1.368721976036E-012\\
Callisto's position & -4.627027927472E-014& -5.146803122579E-013&  4.946080437578E-013\\
Callisto's velocity&  1.145268775648E-013& -4.073677159488E-013&  8.541910668374E-013\\
\botrule
\end{tabular}\label{ipos}}
\end{table}
\begin{table}
  \tbl{Corrections to the initial values of the satellites'
  Keplerian orbital elements.}
 {\begin{tabular}{@{}ccccc@{}} \toprule
Element & Io$\ \ \ \ \ \ \ \ \ \ $ & Europa$\ \ \ \ \ \ \ \ \ \ $ & Ganymede$\ \ \ \ \ \ \ \ \ \ $ & Callisto$\ \ \ \ \ \ \ \ \ \ $\\
\colrule
 $a$ (AU)& -7.846973938685E-013 &  -6.236755903388E-013 &  -4.882674126127E-013 & -3.689635402759E-013\\
 $e$ &  5.625520149536E-010 & 2.509390056859E-010 & -1.085141494890E-010 & -2.457372673403E-011 \\
 $i$ (rad) & 2.998455097495E-011 & -4.1850009742384E-010 & 2.2864355504401E-010 & 1.5703819668256E-010\\
 $\Omega$ (rad) & -3.400988557444E-008 & 1.728156151514E-008 & -1.537507277049E-008 & 3.433878248415E-008\\
 $\omega$ (rad) & 7.4212940681662E-008 & -5.1427109237778E-009 & 1.2778481739417E-007 & -3.4711791663256E-008\\
 $\lambda$ (rad) & -2.520650355108E-010 & 5.146869597183E-010 &  3.610107768281E-010 &  2.891642481017E-011 \\
\botrule
\end{tabular}\label{ikep}}
\end{table}
%
%
\section{Conclusions}

In view of the latest developments in our knowledge of the dynamics of the Jovian system of Galilean satellites, 
in this paper we investigated the influence of the gravitomagnetic field of Jupiter on their motion 
and how difficult a detection of the Lense-Thirring effect would be
still now. We pointed out the presence of
secular drifts on mean longitudes with a highest amplitude of few
tens of kilometers. However, most of this effect will vanish
during the fit procedures that are used in ephemerides
elaboration. Resulting residuals have a maximum magnitude of 20
meters (over 125 years). This appears negligible with today's best
observations which have an accuracy of a few tens of kilometers. A
spacecraft orbiting Jupiter or the adjunction of new more accurate
observations of the Galilean satellites, is definitely required in order to reveal
the Lense-Thirring effect in the Jovian system.

\section*{Acknowledgements}
V.L. benefited from the support of the European Community's
Improving Human Potential Programme under contract
RTN2-2001-00414, MAGE.


\end{document}